\documentclass[review]{elsarticle}

\usepackage{lineno}
\modulolinenumbers[5]
\usepackage{hyperref}

\usepackage[utf8]{inputenc}
\usepackage{amsmath}
\usepackage{amsfonts}
\usepackage{amssymb}
\usepackage{graphicx}
\usepackage{siunitx}
\usepackage{mathrsfs}
\usepackage{multirow}
\usepackage{makecell}
\usepackage{booktabs}
\usepackage{subcaption}
\usepackage[ruled]{algorithm2e}

\journal{Journal of Applied Acoustics}

\DeclareMathOperator*{\argmax}{arg\,max}

\DeclareMathOperator{\STI}{STI}

\newlength\mylen
\newcommand\myinput[1]{%
  \settowidth\mylen{\KwIn{}}%
  \setlength\hangindent{\mylen}%
  \hspace*{\mylen}#1\\}

\graphicspath{ {figures/} }

\begin{document}

\begin{frontmatter}

\title{ Receiver Placement for Speech Enhancement using Sound Propagation Optimization}

\author[unc]{Nicolas Morales\corref{mycorrespondingauthor}}
\cortext[mycorrespondingauthor]{Corresponding author}
\ead{nmorales@cs.unc.edu}
\author[umd]{Zhenyu Tang}
\author[umd]{Dinesh Manocha}
\address[unc]{University of North Carolina at Chapel Hill}
\address[umd]{University of Maryland}


\begin{abstract}
A common problem in acoustic design is the placement of speakers or receivers for public address systems, telecommunications, and home smart speakers or digital personal assistants. We present a novel algorithm to automatically place a speaker or receiver in a room to improve the intelligibility of spoken phrases in a design. Our technique uses a sound propagation optimization formulation to maximize the Speech Transmission Index (STI) by computing an optimal location of the sound receiver. We use an efficient and accurate hybrid sound propagation technique on complex 3D models to compute the Room Impulse Responses (RIR) and evaluate their impact on the STI. The overall algorithm computes a globally optimal position of the receiver that reduces the effects of reverberation and noise over many source positions. We evaluate our algorithm on various indoor 3D models, all showing significant improvement in STI, based on accurate sound propagation.
\end{abstract}


\end{frontmatter}


\section{Introduction}

The acoustic design of a workplace, home, or public venue has a significant influence on the clarity of speech and, consequently, workplace efficiency and comfort. For example, communication problems in a workplace can come from unclear public address systems, unreliable telecommunication devices, or excessive environmental noise. At home, the prevalence of digital personal assistants and control of various home devices via voice recognition commands means that acoustic design can be the difference between a device correctly interpreting a command to turn off an air conditioner or incorrectly interpreting it as one to turn itself off. This class of problems extends to public venues, where the clarity of a speech is dependent on the acoustic design of the venue.

In this paper, we focus on computer-aided design techniques for improving the clarity and intelligibility of speech through optimal speaker or receiver placement. Sound propagates throughout an environment from the source to a receiver and is affected by environmental factors and noise. For example, the effectiveness of teleconference devices in offices is significantly affected by the distance of the user from the device or any obstacles between the user and the device. As such, the acoustic design can be affected by materials, the geometry of the environment, and the placement of sound sources and receivers. Various solutions have been proposed for acoustic material optimization and geometry optimization~\cite{morales2016efficient,monks2000audioptimization}. Some previous work also includes methods for reducing noise in workplace environments~\cite{morales2018optimizing}; however, this paper focuses on receiver placement for the purpose of improving speech intelligibility.

In addition, an interesting trend is the increasing prevalence of speech recognition devices such as Amazon Echo or Google Home~\cite{sriram2018robust}. These devices work by using Automatic Speech Recognition (ASR) algorithms to translate spoken words into text and metadata~\cite{pallett2003look} that can then be processed by the device. Such device can allow certain tasks to be performed more efficiently~\cite{helander2014handbook}. The current trend is to use these devices for organizational purposes, web searching, or control of the home via the Internet of Things (IoT).

In this paper, we use the term \emph{speech recognition} or Automatic Speech Recognition (ASR) to refer to the algorithmic recovery of the spoken source text or metadata from an input recording~\cite{pallett2003look}. These signals can often be noisy or reverberant, making retrieval of the text complicated. In indoor environments, noise can be introduced by secondary sound sources (such as an air conditioning unit, outdoor traffic, or other voices and devices such as televisions), by environmental sound propagation effects such as diffraction and reverberation, and by the characteristics of the receiver microphone. In particular, reverberation can have a detrimental effect on speech recognition algorithms, even when environmental noise is minimized. Its impact on ASR algorithms has been studied extensively~\cite{gillespie2002acoustic}.

The Speech Transmission Index (STI) is a common metric for evaluating the intelligibility of spoken audio~\cite{en1660268}, and is regarded as an accurate subjective measure for human recognition of speech~\cite{galster2007effect}. STI is negatively impacted by reverberation and noise, and thus serves as a useful metric in the evaluation of the intelligibility of a propagation environment and source/receiver configuration. In this paper, we address the problem of computing an optimal placement of the receiver that minimizes error due to sound propagation, environmental effects, and secondary noise sources in order to maximize the STI.

Prior work on ASR techniques has focused on denoising and dereverberation filters on the incoming audio on the receiver in order to reduce extraneous noise~\cite{tashev2005reverberation}. Many of these approaches are based on machine learning techniques for noise minimization~\cite{feng2014speech}, and may need a considerable amount of training data. However, these methods have some limitations. While they can reduce the effects of propagated noise, denoising filters are limited in their applicability to sound propagation paths where the signal to noise ratio (SNR) of the incoming audio is not sufficient to recover the original signal. For example, speech from an adjacent room may only be propagated to the receiver by indirect paths such as propagation through solid walls, diffraction, or reflection. Although previous works have incorporated dereverberation techniques to reduce some of these problems, they often use approximations of reverberation times or decay rates that may not capture the acoustic characteristics of complex environments such as multi-room apartments or offices. This is particularly relevant to how these devices are placed, since solving this kind of design problem can proactively reduce some of these issues. As such, we are interested in placement using computer-aided design techniques rather than the development of real time or training algorithms.

Furthermore, we use STI as a metric for receiver placement quality. First, it is useful as a perceptual metric for human beings. In ASR applications, it has been shown that using features from human auditory systems can improve the performance of speech recognition~\cite{tchorz1999model}. Furthermore, ASR applications are sensitive to noise and STI serves as a perceptual metric for measuring noise from reverberation and secondary sources. Finally, the receiver placement problem can also be posed as a speaker placement problem.

\noindent
\textbf{Main Results:} We present a novel algorithm for receiver placement using sound optimization. Our optimization algorithm maximizes the STI at a receiver by relocating it based on the sound propagation characteristics of an indoor environment. We use a hybrid sound simulation technique for sound propagation, computing the Room Impulse Response (RIR) with wave-based sound simulation techniques at lower frequencies for accuracy and geometric propagation techniques at higher frequencies for performance. We present our optimization algorithm for computing the ideal location for maximizing the STI in Section~\ref{sec:algorithm}. Additionally, using our algorithm, we are able to significantly improve the speech intelligibility at the receiver and minimize the impact of noise and reverberation. We highlight the performance of our algorithm on indoor office and residential scenes (see Section~\ref{sec:results}), where we show a significant improvement of the STI on all scenes.

\section{Prior Work}
\label{sec:prior}

\subsection{Speech Enhancement}
The impact of reverberation and noise remains one of the primary challenges in designing ASR algorithms. There has been extensive work on identifying the impact of various noise sources and on ways of reducing the effect of noise and reverberation on ASR algorithms. Various benchmarks exist to study the effect of noise and reverberation on speech recognition~\cite{hirsch2000aurora}. The CHiME challenge~\cite{barker2015third} aims to promote research on the digital signal processing (DSP) method for noise-robust far field ASR. Setting aside the idea of noise impacts, Gillespie et al.~\cite{gillespie2002acoustic} study the effect of reverberation time ($\text{T}_{60}$) on ASR algorithms. They find that even moderate reverberation causes a catastrophic decrease in the performance of speech-based systems. The REVERB challenge~\cite{kinoshita2016summary} is a novel benchmark specifically designed to test robustness of speech enhancement and ASR techniques for reverberant speech.

Different techniques have been proposed to reduce the impact of reverberation effects on speech recognition. Tashev and Allred~\cite{tashev2005reverberation} use a multi-band decay model for real-time reverberation, but are unable to accurately model all the sound propagation paths. Ko et al.~\cite{ko2017study} use image-source methods for computing an RIR that more accurately captures reverberation, but this is limited to specular reflections. Feng et al.~\cite{feng2014speech} use machine learning techniques to filter out noise and reverberation effects on incoming signals. Palomaki et al.~\cite{palomaki2004binaural} attempt to improve the performance of ASR algorithms by mimicking the binaural properties of human hearing. Chen et al.~\cite{chen2016large} use machine learning techniques to isolate speech features to improve the effectiveness of hearing aids.

The aforementioned benchmarks and techniques are mainly DSP based, some of which incorporate machine learning. Our technique is different and complementary to these methods, and we use an accurate impulse response computation method instead.

\subsection{Acoustic Optimization}
\label{sec:prior_ao}

Previous work in computer-aided design techniques for acoustic optimization has primarily used geometric techniques for sound propagation. Monks et al.~\cite{monks2000audioptimization} use geometric methods to optimize the acoustic materials and shape of room features of acoustic environments. Other material optimization approaches include~\cite{saksela2015optimization} and a 3D wave-based method \cite{morales2016efficient}. Additional wave-based techniques include a 2D FDTD approach~\cite{robinson2014concert} for modifying the shape of balconies in concert halls.

Prior work in speaker placement algorithms include work by Khalilian et al. in sound field reproduction~\cite{khalilian2015joint} and a constraint-based optimization of acoustic treatment and room dimensions for the design of home theater systems~\cite{d1997room}. Our approach is similar to these in that we do not reconfigure the acoustic environment, but rather optimize the placement of the listener.


\section{Evaluating Speech Intelligibility}


\begin{table}[!htbp]
    \centering
    \begin{tabular}{p{1.5cm}p{8cm}}
        $c$                 & the constant speed of sound \\
        $t$                 & time  \\
        $p(\vec{x},t)$      & pressure at location $\vec{x}$ at time $t$ \\
        $p_0$               & reference sound pressure in \si{\pascal} \\
        $h(s,\ell)$         & a room impulse response from source $s$ to listener $\ell$\\
        $H(f)$				& room frequency response
    \end{tabular}
    \caption{Notation and symbols used in our acoustic solver and optimization algorithm.}
    \label{tab:notation}
    \vspace{-10pt}
\end{table}

In this section, we discuss how sound propagates for a given receiver placement and how a receiver placement is evaluated for intelligibility. Notation throughout this section is referenced in Table~\ref{tab:notation}.

\subsection{Speech Transmission Index}

In particular, we are interested in the Speech Transmission Index (STI). STI is an objective measure to evaluate speech intelligibility of a transmission channel that has been widely applied since 1970s \cite{houtgast2002past}. It provides reliable results that agree with subjective measures \cite{wijngaarden1999objective}, and is effective for different languages \cite{houtgast1984multi}. Thus, we use STI as our objective in optimization.

The basis of STI is the computation of the Modulation Transfer Function (MTF) \cite{houtgast1985review}. 
In our work, we adapt the indirect method where the simulated RIR is used to derive MTF \cite{schroeder1981modulation} of the transmission channel, using equation~(\ref{eq:MTF}):
\begin{equation}
\label{eq:MTF}
m_k(f_m)=\frac{|\int_{0}^{\infty}h_k(t)^2e^{-j2\pi f_mt}dt|}{\int_{0}^{\infty}h_k(t)^2dt}\times (1+10^{-\text{SNR}_k/10})^{-1},
\end{equation}

\noindent where $r_k(t)$ is our impulse response filtered to octave band $k$, $f_m$ is the modulation frequency defined in \cite{en1660268}, and $\text{SNR}_k$ is the SNR in octave band $k$ in decibels, which is further explained in Section~\ref{sec:secondary_noise}. Note that the noise here is a combination of physical noise, threshold and masking effects, which is not provided with the RIR, but can be simulated with our propagation model. Consequently, both environment noise and reverberation negatively impact the STI. The result $m_k(f_m)$ is the modulation transfer ratio at modulation frequency $f_m$. STI can be calculated from a weighted contribution of $m_k(f_m)$.
Due to sound frequency's dependency on gender, STI has a separate set of frequency band weightings for males and females. Without losing the generality of our result~\cite{cabrera2018critical}, we use male weightings throughout our optimization. For robust implementation of STI from RIR, we refer to \cite{cabrera2014increasing}.

\subsection{Hybrid Sound Propagation}
\label{sec:hybrid}

The RIR is used to model the sound propagation effects in a room for given source and receiver positions. Given an impulse sound (e.g. one similar to a Dirac delta function) at a source location, the sound pressure is evaluated at the receiver to determine the RIR. Traditional geometric approaches for sound propagation rely on the assumption that sound travels geometrically rather than as a wave; as a result, such approaches do not provide accurate representations of certain wave phenomena like diffraction and scattering. Wave-based methods, on the other hand, rely on numerical solvers of the acoustic wave equation:

\begin{equation}
\label{eq:wave}
\frac{\partial^2}{\partial t^2} p(\vec{x},t)-c^2\nabla^2p(\vec{x},t)=f(\vec{x},t).
\end{equation}

However, the computational cost of wave-based solvers becomes intractable at higher frequencies. The asymptotic complexity of these methods scales with $O\left(f^4\right)$ where $f$ is the simulation frequency. Thus, wave-based methods are computationally expensive at higher frequencies, but can accurately represent wave phenomena prevalent at lower frequencies. This is important because human speech often includes a wide range of frequencies and much of the energy is concentrated at lower frequencies.

In optimization techniques, it is often the case that the objective function must be evaluated many times. In order to maintain both accuracy and efficiency, we evaluate the RIR and subsequently the STI metric through the use of a hybrid sound propagation scheme. We combine impulse responses from a wave-based method and a geometric method for sound propagation. Adaptive Rectangular Decomposition (ARD)~\cite{raghuvanshi2009efficient, morales2015parallel}, an efficient solver for indoor scenes, is used for computing the lower frequencies (up to \SI{500}{\hertz}) of the impulse response. We use a ray tracing solver~\cite{schissler2016interactive} that simulates both specular and diffuse reflections for higher frequencies. The wave-based method is more accurate for lower frequencies, where diffraction and scattering effects are more apparent, while the ray-tracing method is more computationally efficient.



Given a frequency response $H_w(f)$ computed by the wave-based solver up to \SI{500}{\hertz} and a frequency response $H_g(f)$ for the geometric technique, we can determine the hybrid impulse response with the application of a Linkwitz-Riley crossover filter~\cite{linkwitz1976active}:

\vspace{-1em}
\begin{equation}
\label{eq:lr4}
h(t)=\mathscr{F}^{-1} \left\{B^2_{low}( H_w(f) ) + B^2_{high}( H_g(f) )\right\},
\end{equation}

\noindent where $B^2_{low}$ is the composition of two Butterworth lowpass filters and $B^2_{high}$ is the composition of two Butterworth highpass filters. The application of the Linkwitz-Riley crossover filter helps avoid ringing artifacts from the lowpass and highpass stages.

In our algorithm, we use pre-recorded sound clips of a human voice for the propagated sound. We use the convolution of the sound clip with the impulse response to yield the propagated sound:

\begin{equation}
    I(s_i,\ell) = h(s_i,\ell) \circledast b_i,
\end{equation}

\noindent where $b_i$ is the sound clip associated with the source.

\subsection{Environmental Noise}
\label{sec:secondary_noise}
One important aspect of speech intelligibility is the ambient or environmental noise present in the domain. In addition to evaluating primary sound from speaker positions in our objective function, we simulate noise emitted from secondary sources, such as a television or HVAC system. The secondary noise is then propagated separately by our hybrid solver and used in computing the SNR for the MTF:

\begin{equation}
    \text{SNR}_k = 10\log_{10}\left(\frac{I(s_\text{secondary}, \ell)_k + I_{\text{masking},k} + I_{\text{threshold},k}}{I(s_i, \ell)_k}\right),
\end{equation}

\noindent where $I_\text{threshold,k}$ is the auditory reception threshold and $I_\text{masking,k}$ is the auditory masking from the combined noise from the primary and secondary sources~\cite{cabrera2014increasing} for the $k$-th octave band.

\begin{figure*}[!htp]
    \centering
    \begin{subfigure}[b]{0.32\linewidth}
        \centering
        \includegraphics[width=\linewidth]{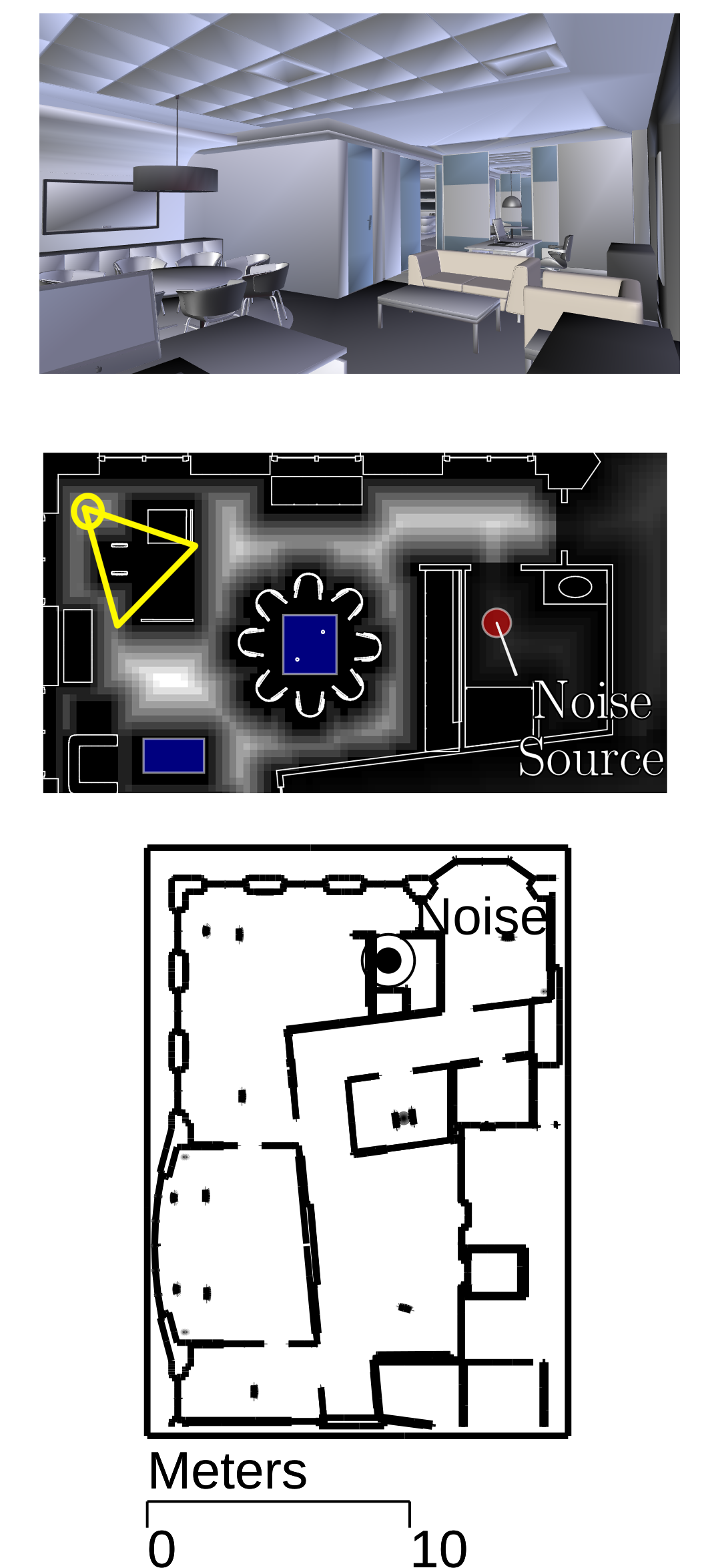}
        \label{fig:office_view}
        \vspace{-2em}
        \caption{Office (Zoomed-in)}
    \end{subfigure}
    \begin{subfigure}[b]{0.32\linewidth}
        \centering
        \includegraphics[width=\linewidth]{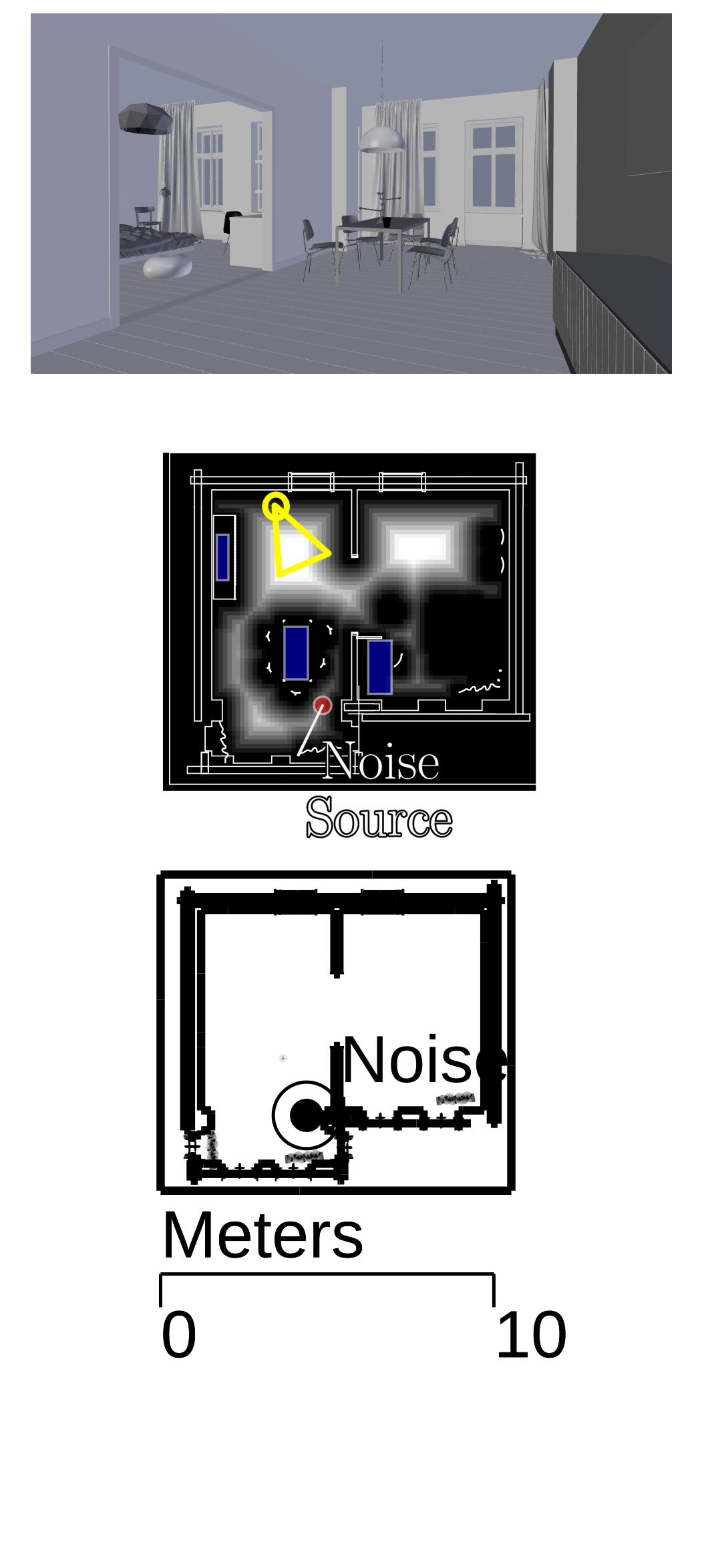}
        \label{fig:berlin_view}
        \vspace{-2em}
        \caption{Berlin}
    \end{subfigure}
    \begin{subfigure}[b]{0.32\linewidth}
        \centering
        \includegraphics[width=\linewidth]{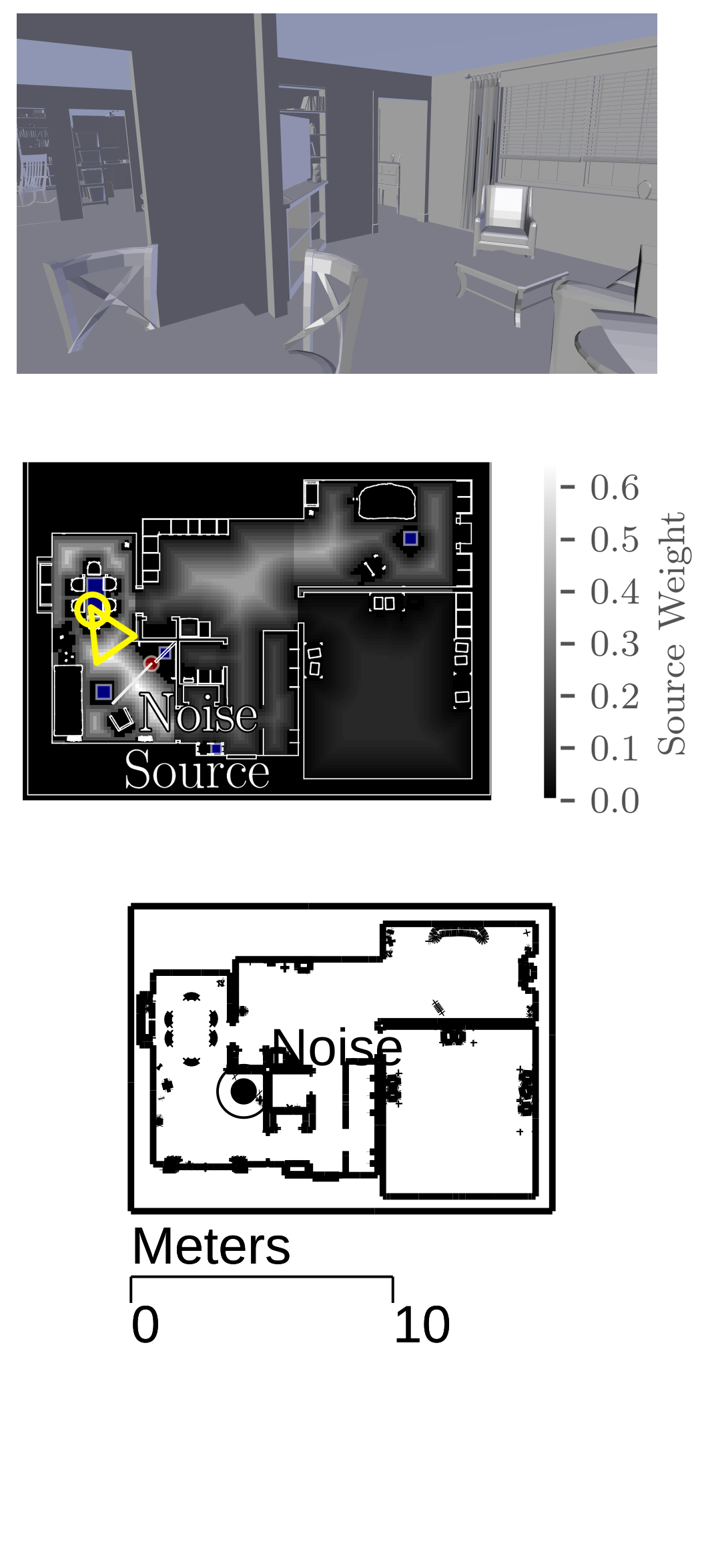}
        \label{fig:suburban_view}
        \vspace{-2em}
        \caption{Suburban}
    \end{subfigure}
    \vspace{-1em}
    \caption{The three complex CAD benchmarks used to evaluate our algorithm: The Office scene, the Berlin scene, and the Suburban scene. The first row shows the interior of part of the CAD scene, with the camera positions marked in the second row in yellow. The second row shows diagram of the placements constraints used in the discretization of a portion of these scenes from a top-down view. Blue regions correspond to allowable areas within which the listener can be placed. The gradient represents possible source positions corresponding to where a human speaker may be located and the weighting of that speaker location. The red dot refers to a noise source that can interfere with STI. Note that subfigure (a) is a zoomed-in view of the office scene, in the area of interest with positive weights since the rest of the benchmark has very low weights. The third row shows the dimensions of the rooms and locations of the noise sources.}
    \vspace{-1em}
\label{fig:setup}
\end{figure*}

\section{Our Optimization Algorithm}
\label{sec:algorithm}
In a typical environment where speech recognition is used, having a high STI for a single source location is not sufficiently optimal for the entire environment. For example, in a household, the user of a speech recognition device could use the device from many different locations. As the user's location changes, so does the source position and the propagated sound. Therefore, we consider the set of all possible source locations $S$ in our optimization formulation, based on user-defined constraints. This set is sampled discretely according to a uniform distribution yielding the source locations $s_1 \dots s_n$ where $n$ is the number of sampled locations.

\subsection{Objective Function}

\begin{figure}[!htbp]
    \centering
    \includegraphics[width=\linewidth]{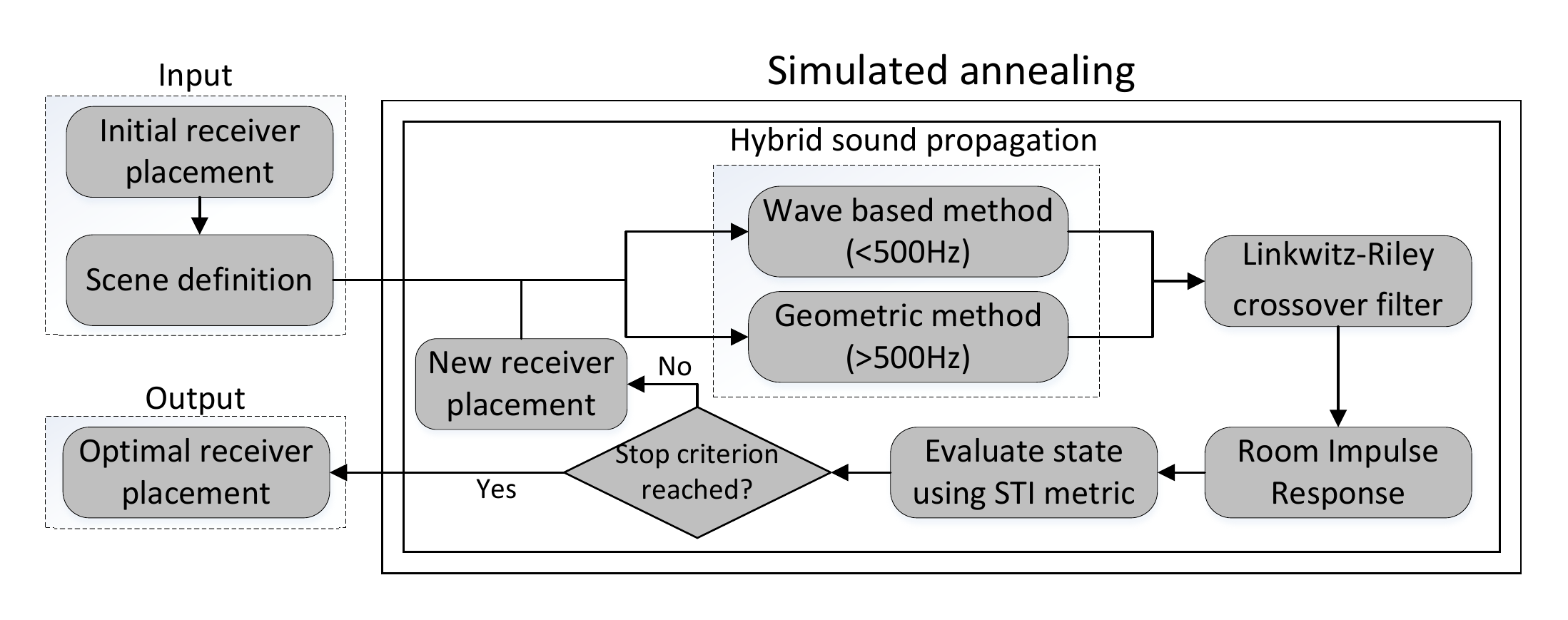}
    \vspace{-2em}
    \caption{ \textbf{We highlight various components of our approach:} The scene definition consists of the scene geometry (i.e. the triangulated CAD model) and the acoustic property of each material assigned to the triangular mesh. It is sent to our optimization scheme as the input along with an initial receiver location. The RIR is computed using a hybrid sound propagation approach, with a wave based technique under \SI{500}{\hertz} and a geometric technique above \SI{500}{\hertz}. The STI is then computed from the RIR and averaged to evaluate the objective function. Our simulated annealing approach then computes a new receiver location for the next iteration until the stopping criteria are encountered.}
    \vspace{-1em}
\label{fig:overview}
\end{figure}

Given the optimization variable for the receiver location $\ell$, we use the following objective function:

\begin{equation}
\label{eq:objective}
\argmax_\ell \sum_{i=1}^n w_i \STI\left( h\left( s_i, \ell \right) \right),
\end{equation}

\noindent where $w_i$ is a weighting for the source location $s_i$ defined by the user. An overview of how the objective function is used to drive the overall optimization of the receiver location is shown in Figure~\ref{fig:overview}.

The goal of this objective function is to find the receiver location where the STI is maximized throughout the domain. Importantly, this depends on the acoustic environment and how sound propagates in it. The STI is computed using the hybrid impulse response described in Equation~\ref{eq:lr4}. The linear weighted sum allows for some designer control in the multiobjective optimization process.

Using this weight, certain regions of the environment can be prominent or dominant in our optimization algorithm. For example, if a speech recognition device is primarily for use in the living room, source locations in that room should be weighted higher. This weighting is specified by the designer, but can also be computed using data-driven approaches; for example the measured frequency at which the user is in each room or the areas in which workers are primarily located.

A higher weight in a particular location will make the optimization process more sensitive to changes of the STI in that area. For example, in our Suburban scene, we weight the garage area close to zero. As a result, the STI in the garage has little effect on the results of the optimization process. However, both living rooms were weighted highly, leading the algorithm to select a location in one living room, but still be intelligible in parts of the other.

To compute the objective function for a specific listener position, we compute the RIR and STI for every source. In our implementation, we propagate acoustic waves outwards from the receiver position rather than the source in order to evaluate $h\left(s_i,\ell\right)$ for all source locations $s_1 \dots s_n$ with the wave-based technique. In general, the cost of the ARD method is the same whether one listener is evaluated or the entire field is computed. Therefore, using acoustic reciprocity the same holds for whether one source is active or a large number of sources are active for one listener. If there is a many-to-one or one-to-many relationship between sources and listeners, it is possible to take advantage of this property.

Without loss of generality, a similar approach without reciprocity can be used to determine speaker placement in a public address system, though our experiments focus on receiver placement.

\subsection{The Optimization Domain}

The optimization domain is first defined in continuous space by the set of subdomains that are distinct regions of space. In our implementation, these are defined by the designer using 3D axis-aligned bounding boxes, although it is straightforward to use other subdomain shapes. Each of these subdomains represents constraints on where the receiver can be placed. This can reflect structural constraints, such as the placement of a PA system, or aesthetic constraints such as the location of a home automation device.

This domain is then uniformly sampled, using a stratified sampling approach. Our set of listeners is defined as:

\begin{equation}
L=\left\{\ell:\ell \in \bigcup\limits_{i=1}^m{B_i}\right\},
\end{equation}

\noindent where $L$ is the set of possible listeners and $B_i$ is a subdomain specified by the designer.

To maximize Equation~(\ref{eq:objective}), we select from a set of discrete receiver locations $L$. This discrete sampling allows for a straightforward way of evaluating constraints. The primary constraint of the receiver location is an allowable set of surfaces on which the receiver can be placed. A device would commonly be placed on a table or counter top, but the floor would not be a desirable location. Receiver locations are sampled from the areas allowed by the constraints. Figure~\ref{fig:setup} shows these constraints on a typical scene, such as an office workplace.

The choice of sampling density and distribution can affect the convergence of the optimization algorithm and the performance of the approach. For example, too coarse a sampling can cause some details of the resulting pressure field to be missed, while too fine a sampling can increase the overall search space of the optimization algorithm and increase the overall number of iterations.

In our implementation, we used a sampling density of one sample every approximately \SIrange{5}{10}{\centi\meter}. The spatial step of the wave-based solver for \SI{500}{\hertz} was approximately \SI{10}{\centi\meter}, so this sampling was able to sufficiently capture variances in the lower-frequency sound field.

\subsection{Simulated Annealing}

 \begin{algorithm}[t]
    \SetAlgoLined
 	\KwIn{Source regions $\Omega$, $s_1,...,s_n\in \Omega$}
 	\myinput{Initial temperature $T_0$}
 	\myinput{Cooling rate $\alpha$}
 	\KwOut{Optimal receiver location $l_{opt}$}
 	\SetKwFunction{Initializaiton}{Initializaiton}
     \SetKwFunction{PermuteState}{PermuteState}
     \SetKwFunction{TestState}{TestState}
     \SetKwFunction{GetInitialState}{GetInitialState}
    
     \Initializaiton{ }\;
     $l$ $\leftarrow$ \GetInitialState{ }\;
     $q \leftarrow \sum_{i=1}^n w_i \STI\left( h\left( s_i, \ell \right) \right)$\;
     $T \leftarrow T_0$\;
     \While{$T>1$}{
         $l' \leftarrow$ \PermuteState{l};  /* Compute new state */\\
         $q' \leftarrow \sum_{i=1}^n w_i \STI\left( h\left( s_i, \ell \right) \right)$\;
         \If{\TestState{$q,q',T$}}{
             $l\leftarrow l'$\;
             \If{$q'>q$ }{
                 $q\leftarrow q'$;\quad/* Update optimal state */\\
                 $l_{opt}\leftarrow l'$\;
             }
         }
         $T\leftarrow \alpha T$;\quad/* Temperature cools down */\\
     }
     \SetKwProg{myproc}{Procedure}{}{}
     \SetKwFunction{Rand}{Rand}
     \myproc{\TestState{$q,q',T$}}{
         \If{$q' > q$}{
             \KwRet{true};  /* Always accept better state */\\
         }
         $p\leftarrow e^{\frac{q-q'}{T}}$; /* Accept worse state by probability */\\
         \KwRet{$p < \Rand{0,1}$}\;
     }
 	\caption{Simulated Annealing}
 	\label{alg:SA_STI}
\end{algorithm}

Given the discrete search domain $L$, we use a simulated annealing (SA) approach for choosing the optimal receiver position and maximizing STI. SA algorithms work by mimicking physical annealing processes, where gradual cooling processes affect the structure of a material. In SA, an optimal solution to an optimization problem is obtained by randomly selecting configurations and gradually restricting (or \emph{cooling}) the choice of a new configuration. 

The probabilistic nature of SA techniques allow them to avoid local extrema by probabilistically iterating over less-optimal solution. Early termination conditions allow our algorithm to keep the number of iterations low. Additionally, since the compute cost of evaluating the acoustic field in our hybrid approach is very high, simulated annealing improves iteration performance by only evaluating the field once for each iteration.

In typical SA approaches, an initial temperature $T_0$ is chosen. Then, using a specified cooling schedule, the temperature is reduced on each iteration as the optimization variables are perturbed. The temperature is used to determine the probability of moving to a less-optimal state, where the optimality is given by the energy of that state. Whenever a new state is chosen for the optimization variables, better states are allowed but worse states may still be permitted depending on the temperature. In our formulation, given a configuration of listener location $\ell$, the energy is the STI computed in  Equation~\ref{eq:objective}. Since states represent possible receiver positions, a new state $\ell^\prime$ is determined by randomly selecting a new discrete point from the set of possible receiver positions. The entire SA algorithm is described in the pseudo-code listing Algorithm~\ref{alg:SA_STI}

We tune the cooling schedule of the SA algorithm so that at the end of our optimization process, the probability of accepting a state with an STI $0.03$ (the just-noticeable difference of STI~\cite{bradley1999just}) less than the current state is $1\%$.

\begin{equation}
T_{end}=-\frac{0.03}{\ln{0.01}}.
\end{equation}

Additionally, we reduce the overall number of iterations by ending early if a state is rejected $k$ times, where $k=10$ yielded good results in our experiments. Usually a long sequence of consecutive rejects means that the SA approach has encountered a global maximum, or at least that the probability of a better configuration is low.

\subsection{Comparison to Prior Work}

In Section~\ref{sec:prior_ao}, we discuss prior methods of acoustic optimization dealing with various acoustic problems. Monks et al.~\cite{monks2000audioptimization} also use a simulated annealing technique, but for shape and material optimization. However, the technique uses geometric methods that are not accurate for lower frequencies. Similarly, prior work in wave-based optimization for materials~\cite{morales2016efficient} is restricted to lower frequencies where the computational cost of the wave-based method is not intractable.

Recent work has focused on hybrid acoustics for noise control~\cite{morales2018optimizing}. Our problem formulation differs in a few aspects. First, we are interested in a problem domain with fixed environmental noise, and rather the improvement of STI by the placement of speakers or receivers. Our technique can work in conjunction with previous work --- consider a design pipeline where noise in minimized using prior work and the result of the minimization is used as the static noise input for improving speech intelligibility. Secondly, since we are dealing with a single omnidirectional receiver/listener, we can take advantage of some properties of acoustic wave propagation to improve the performance of our algorithm; using the principle of acoustic reciprocity, we can model receivers in the same way we model sources. This is only possible with a one-to-many relationship between receivers and sources rather than a many-to-many relationship. Finally, we focus on different acoustic phenomena. While environmental noise is an important focus of our work, we are also interested in the reverberation in the room where the listener is being placed. Therefore optimal placement is more heavily dependent on the acoustic environment, not just the secondary noise sources.

\section{Results and Analysis}
\label{sec:results}
\begin{table}
    \centering
    \resizebox{\linewidth}{!}{
    \begin{tabular}{@{}ccccccccccccc@{}}
        \toprule
        STI & $>$0.76 & 0.74 & 0.7& 0.66& 0.62 & 0.58 & 0.54 & 0.5 & 0.46 & 0.42 & 0.38 & $<$0.36
        \\ \hline 
        Quality rating & A+ & A& B&C&D&E&F&G&H&I&J&U
        \\ \bottomrule
    \end{tabular}}
    \caption{STI scale and qualification~\cite{en1660268}}
    \label{tab:sti_quality}
    \vspace{-1em}
\end{table}

\begin{table*}
    \centering
    \begin{tabular}{@{}lrrrrrrr@{}}
        \toprule
        \thead{Scene} & \thead{Num.\\Triangles} & \thead{Model Size} & \thead{Avg.\\STI Before} & \thead{Avg.\\STI After}  & \thead{Num.\\Iterations} & \thead{Num.\\Samples} & \thead{Time (\si{\second})}
        \\ \hline
        Office & 973373 & $1290m^3$ & $0.1600$ & $0.6757$ & $60$ & $879$ & $2357$
        \\
        Berlin & 2198393 & $370m^3$ & $0.1818$ & $0.5601$ & $76$ & $219$ & $2707$
        \\
        Suburban & 85604 & $391m^3$ & $0.1802$ & $0.6571$ & $37$ & $554$ & $1686$
        \\ \bottomrule
    \end{tabular}
    \caption{Our optimization process can improve the STI of scenes of differing complexity. We were able to improve the receiver's STI in the Berlin scene from the lowest intelligibility rating of U to an intelligibility rating of E, which is suitable for high quality public address systems. Our optimization process improved intelligibility of the suburban scene and the office scene from U to C, which indicates high speech intelligibility (see Table~\ref{tab:sti_quality}). We also reference the relative scene complexity for both the wave-based and geometric method. Since the performance of the geometric method is dependent on the triangle mesh for the surface representation of the model, we show the total number of surface triangles in each benchmark. We also summarize the volume, since the wave-based method uses a spatial regular grid discretization of the input model. The number of samples is the number of discrete listener positions and the size of the search space.}
    \vspace{-1em}
    \label{tab:performance}
\end{table*}

We tested our receiver placement optimization on three workplace and residential scenes. Our method can accurately compute the STI on complex indoor scenes. The optimization was computed on a desktop machine, using 8 threads. Our benchmark scenes included Office, a multi-room workplace with conference rooms; Suburban, the ground floor of a multi-story house; and Berlin, a small apartment with two connected rooms. The material absorptions of the three benchmarks are specified by hand annotation of their respective CAD models using the material database from~\cite{egan2007architectural}.

We evaluate the performance of our algorithm on different benchmarks. On each scene we were able to obtain a result significantly greater than the JND of STI, which is 0.03~\cite{bradley1999just}. A reference for the intelligibility of different STI values is included in Table~\ref{tab:sti_quality}. Our optimization algorithm is able to converge in a few iterations to the maximum STI. These results are summarized in Table~\ref{tab:performance}. In this table, we also highlight the complexity of the scene with respect to the volume (which affects the performance of the wave-based model) and with respect to the number of triangles in the CAD model (since the geometric solver works by tracing rays against a triangle mesh).

\begin{figure}[!htbp]
    \centering
    \begin{subfigure}[b]{\linewidth}
        \centering
        \includegraphics[width=0.7\linewidth]{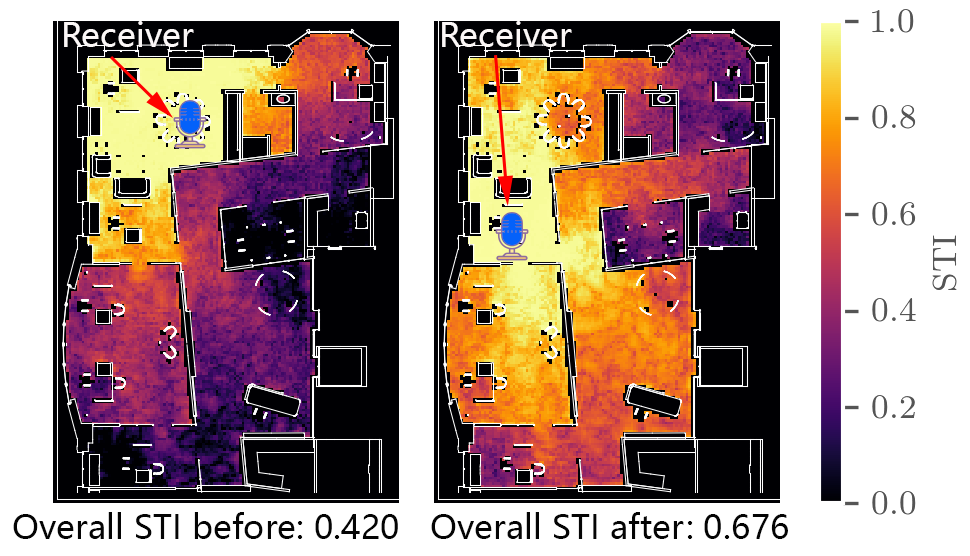}
        \label{fig:office_field}
        \caption{Office scene STI fields}
    \end{subfigure}
    \begin{subfigure}[b]{\linewidth}
        \centering
        \includegraphics[width=0.75\linewidth]{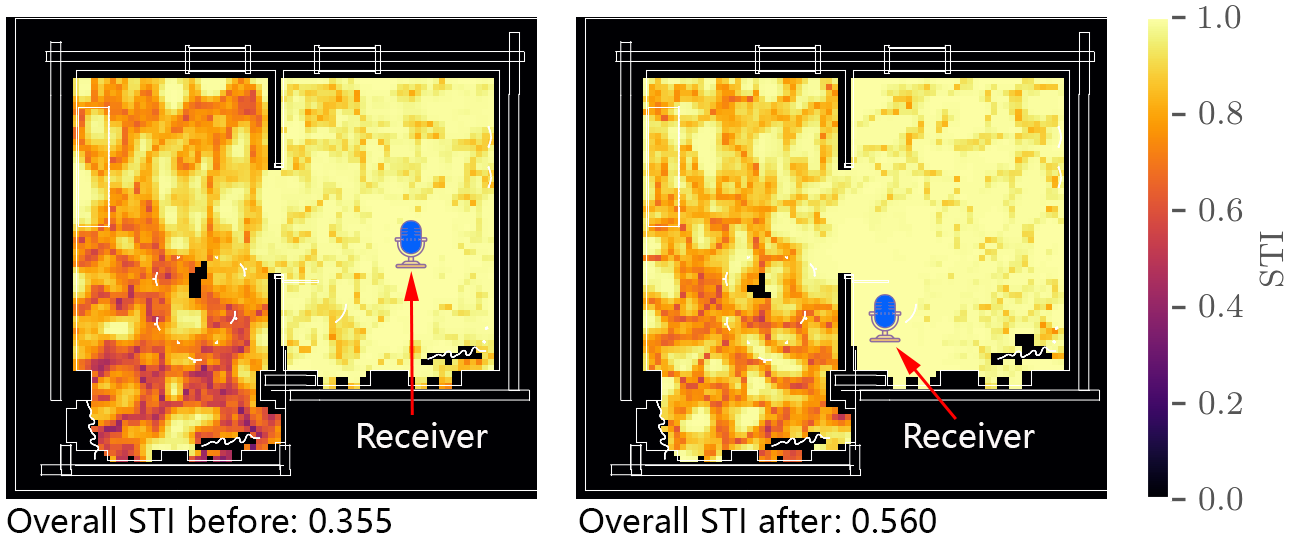}
        \label{fig:berlin_field}
        \caption{Berlin scene STI fields}
    \end{subfigure}
    \begin{subfigure}[b]{\linewidth}
        \centering
        \includegraphics[width=0.85\linewidth]{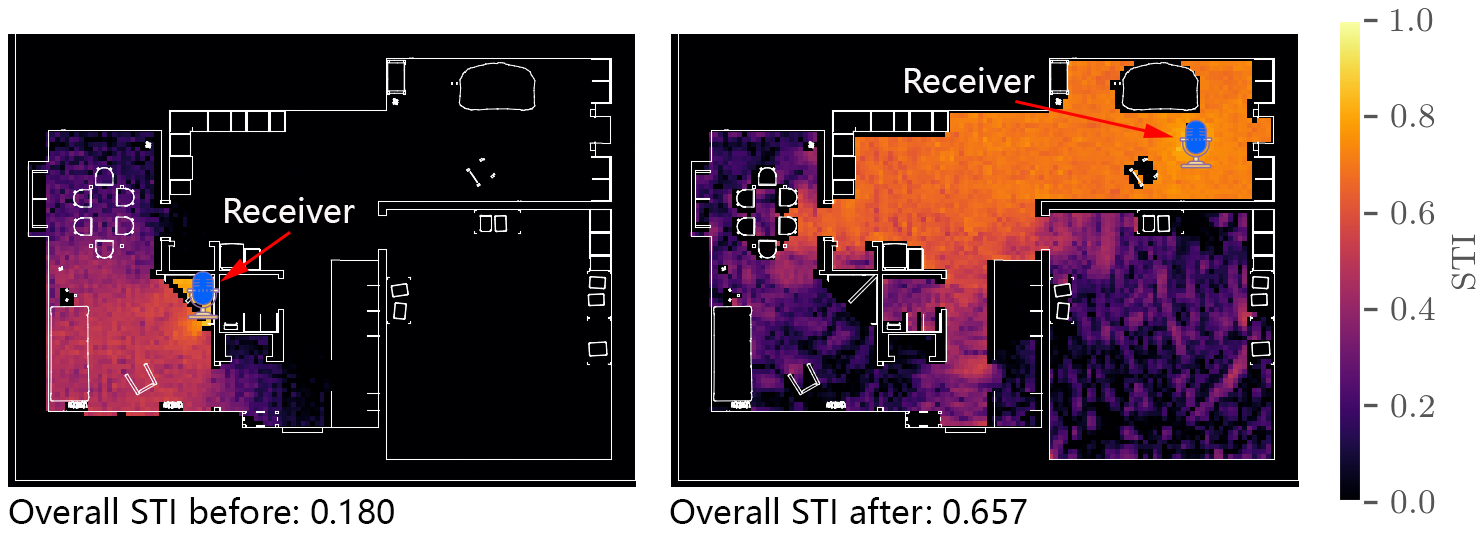}
        \label{fig:suburban_field}
        \caption{Suburban scene STI fields}
    \end{subfigure}
    \caption{Side-by-side comparison of the initial and final placements for the receiver according to our optimization process. The figure on the left shows a receiver placement close to the noise-emitting source, yielding a very poor STI rating. The figure on the right shows placement as a result of our optimization process. The receivers are placed away from the noise source, but also placed to avoid areas that have reflective floors. Note that the left side placements are generally not the worst case because we do not search the whole feasible areas by brute force. Consequently, the actual improvement on STI could be greater in worse cases.}
    \vspace{-1em}
\label{fig:fullfield_sti}
\end{figure}

\begin{figure}[h]
    \centering
    \begin{subfigure}[b]{\linewidth}
        \centering
        \includegraphics[width=0.5\linewidth]{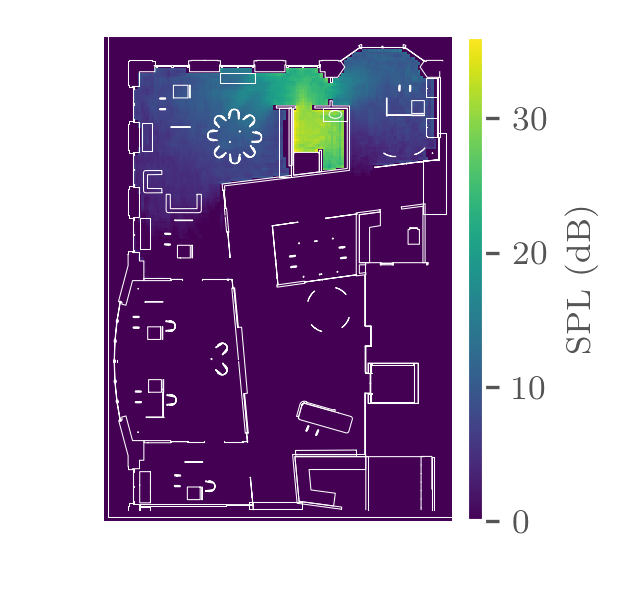}
        \label{fig:office_pressure}
        \caption{Office scene noise pressure distribution}
    \end{subfigure}
    \begin{subfigure}[b]{\linewidth}
        \centering
        \includegraphics[width=0.5\linewidth]{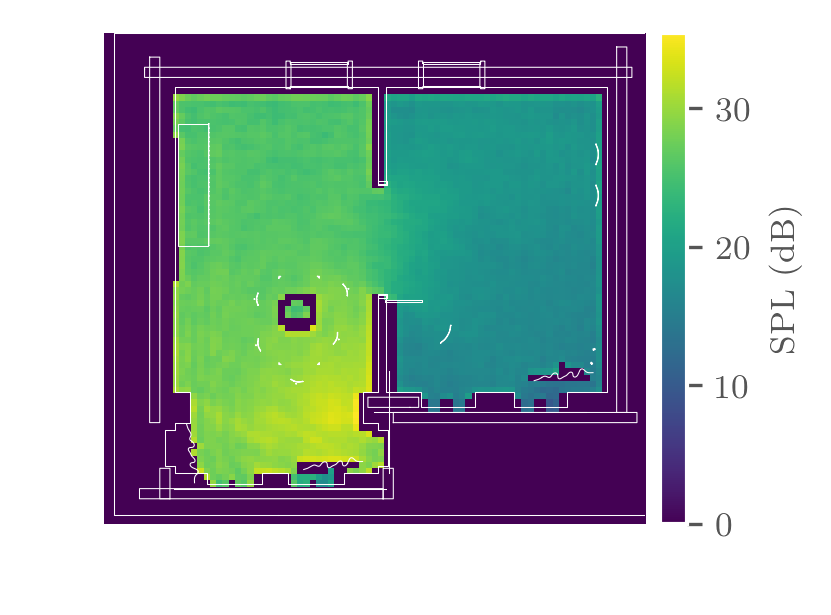}
        \label{fig:berlin_pressure}
        \caption{Berlin scene noise pressure distribution}
    \end{subfigure}
    \begin{subfigure}[b]{\linewidth}
        \centering
        \includegraphics[width=0.5\linewidth]{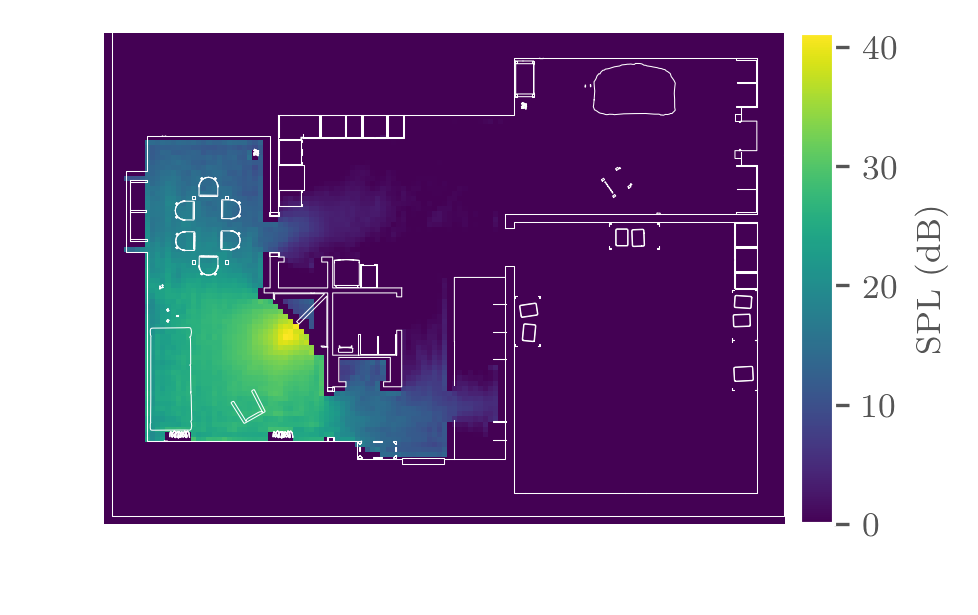}
        \label{fig:suburban_pressure}
        \caption{Suburban scene noise pressure distribution}
    \end{subfigure}
    \caption{Pressure distribution of each of the noise sources in the three models.}
\label{fig:pressure_dist}
\end{figure}

Figure~\ref{fig:fullfield_sti} shows the impact the choice of listener position has on the various speaking positions within the Suburban scene. Our optimization process accounts for multiple speaking locations throughout the environment rather than a single location. The pressure distribution from the noise source in each scene is summarized in Figure~\ref{fig:pressure_dist}.

\begin{figure}[htbp]
    \centering
    \includegraphics[width=0.8\linewidth]{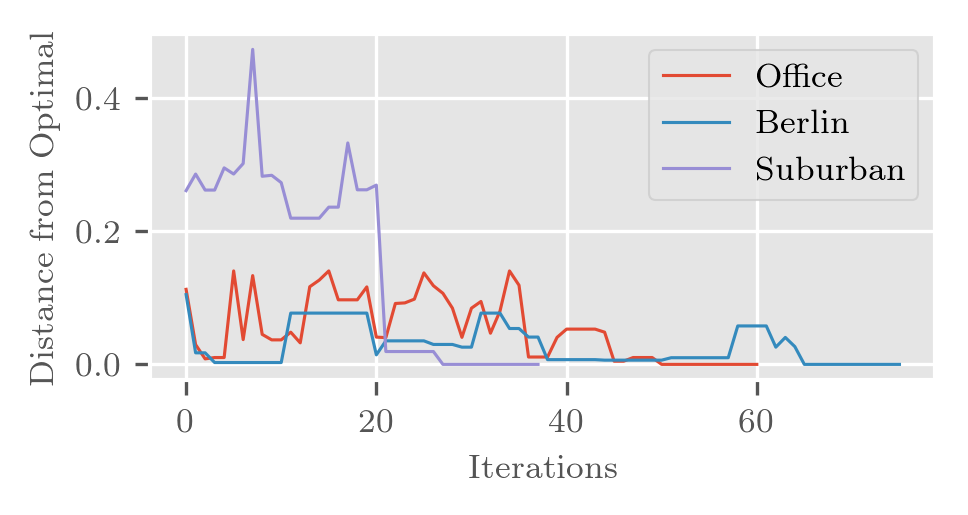}
    \vspace{-1em}
    \caption{Convergence plot of our optimization process in different environments. These show the overall energy difference at each iteration in comparison to the final converged energy. The starting point was selected randomly. The distance to optimal is a measure of the absolute value difference between the total energy in each iteration and the final energy at the end of the optimization process.}
    \label{fig:convergence}
    \vspace{-1em}
\end{figure}

A summary of the convergence of our algorithm is in Figure~\ref{fig:convergence}. The starting points for these experiments were selected randomly, and show how the simulated annealing process avoids local maxima throughout the optimization.

\subsection{Validation}

\begin{table}
    \centering
    \begin{tabular}{@{}lccccc@{}}
        \toprule
        & \thead{Measured} & \thead{Geometric} & \thead{Wave-based} & \thead{\textbf{Hybrid}} & \thead{\textbf{Hybrid Error}}
        \\ \hline 
        IR0 & $0.8438$ & $0.7186$ & $0.7241$ & $\mathbf{0.7510}$ & $\mathbf{11\%}$ \\
        IR1 & $0.7251$ & $0.6838$ & $0.6885$ & $\mathbf{0.7031}$ & $\mathbf{3\%}$
        \\ \bottomrule
    \end{tabular}
    \caption{Comparison of our simulated STI with measured STI in a digitally scanned room by~\cite{schissler2018acoustic}.}
    \label{tab:ir_validation}
\end{table}

We measured the accuracy of our hybrid propagation STI calculations by comparing our method to measured IRs in a room that was digitally scanned in work by~\cite{schissler2018acoustic}. These results are summarized in Table~\ref{tab:ir_validation}. For the comparison, we cutoff the simulated IR to match the total length (in samples) of the measured IR and matched the peak offset to account for differing impulse times in the measured and simulated impulse responses.

Previous work on both the geometric and wave-based approaches we used has also performed some validation work. Schissler et al.~\cite{schissler2017interactive} compare the accuracy of their geometric method (which we use) on the round Robin Elmia benchmark. Additionally the accuracy of the ARD wave-based solver was evaluated with measured IRs in~\cite{mehra2014acoustic}.

\subsection{Comparison to STI estimation}

\begin{figure}[!htbp]
    \centering
    \includegraphics[width=0.75\linewidth]{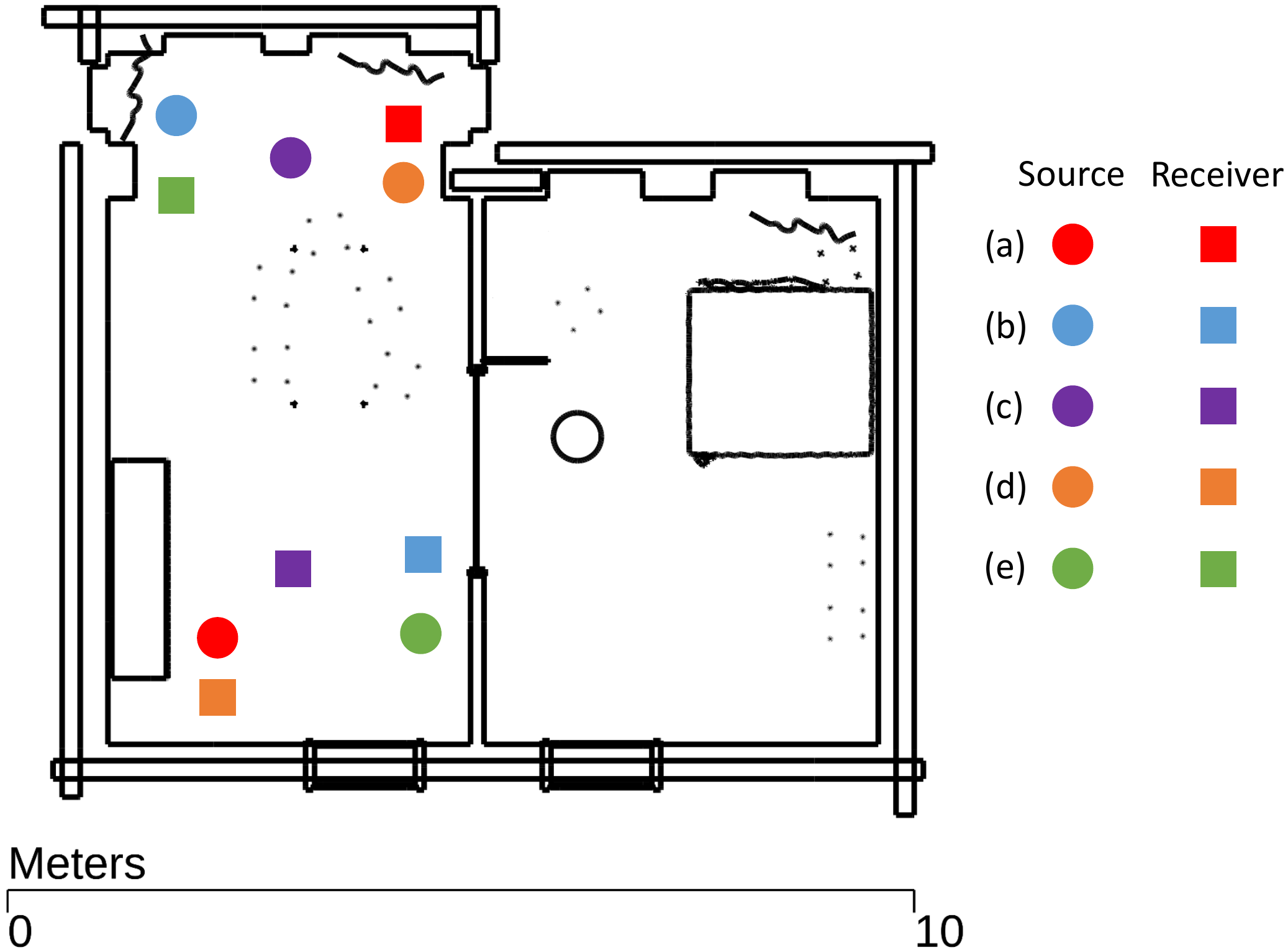}
    \caption{Experiment setup for evaluating effects of inaccurate propagation. Five source-receiver pairs labeled from ($a$) to ($e$) are manually chosen throughout the Berlin scene. Note that we separated out a single room by adding a wall for the convenience of applying Sabine approximation.}
    \vspace{-1em}
    \label{fig:sabine_setup}
\end{figure}
\begin{figure}[!htbp]
    \centering
    \includegraphics[width=0.82\linewidth]{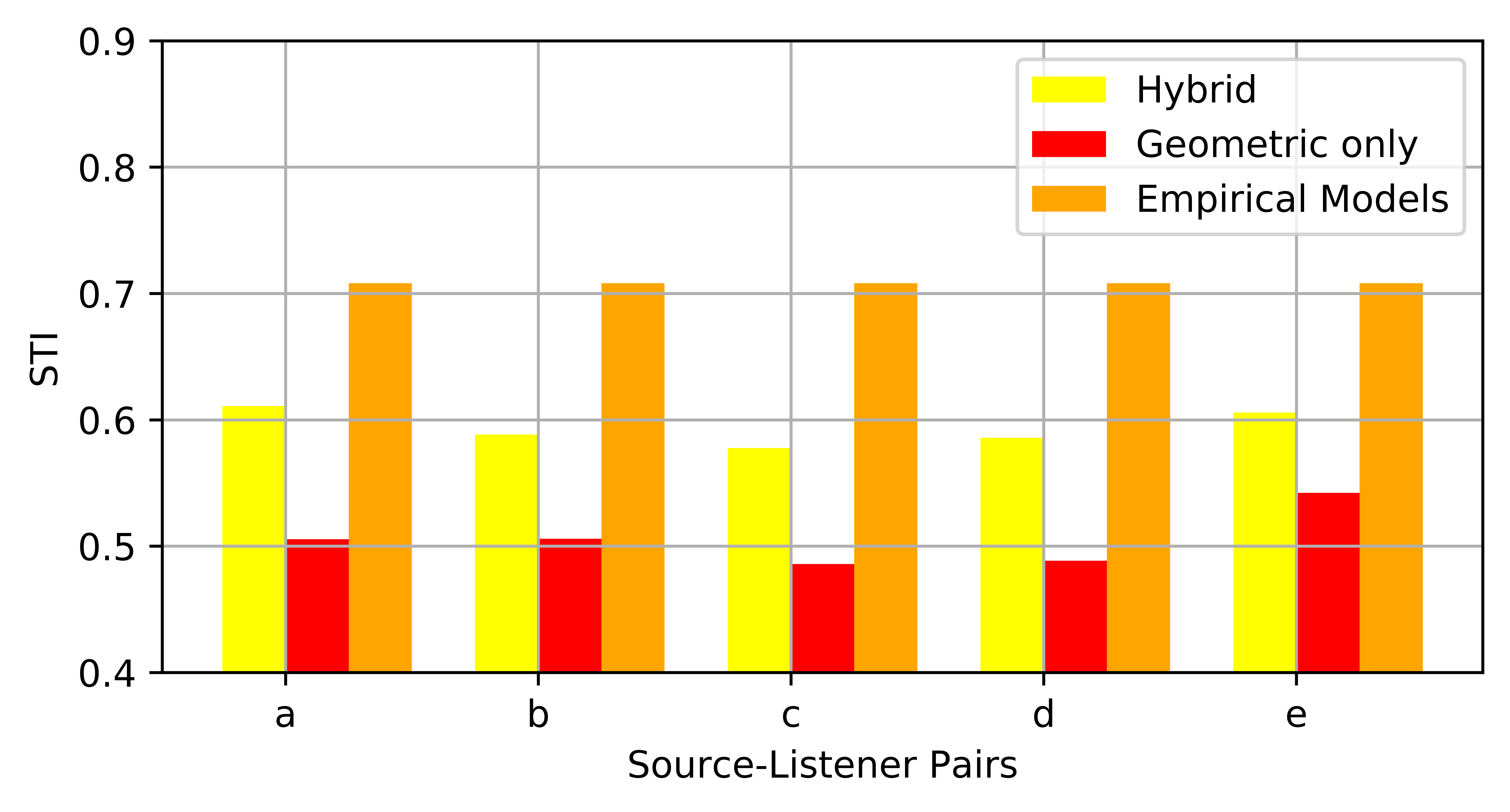}
    \caption{Comparison of the accurate hybrid sound propagation model to using less accurate propagation models. Source-listener pairs \emph{a} through \emph{e} were arbitrarily selected from the Berlin scene. The \emph{hybrid} and \emph{geometric only} STI values displayed are calculated from impulse responses generated by a full hybrid sound propagation, and a simulation using only geometric sound propagation respectively; and the \emph{empirical} STI is calculated from empirical models in \cite{diaz2005reverberation,tang2004speech}, which do not account for source/listener location. We observe that the difference between the hybrid model and the other two is always above the JND of STI. It is not sufficient to use less accurate propagation models even for simpler scenes such as the Berlin scene.}
    \vspace{-1em}
    \label{fig:sabine}
\end{figure}

Additionally, we compare our results to the use of less accurate reverberation models, including empirical predictive models for reverberation time~\cite{diaz2005reverberation} and STI~\cite{tang2004speech} based on room volume, and the geometric propagation model that is part of our hybrid simulator. Previous research~\cite{morales2018optimizing} has shown that a pure geometric approach can lead to errors up to 36\si{\decibel} in frequency response at a low frequency (\SI{125}{\hertz}) compared to wave-based methods in an area of diffraction. Here we evaluate the error introduced to STI from inaccurate sound propagation. We show errors beyond the JND of STI that occur while using these models. The single room we used for the test has a dimension of $4.475\times8.338\times 3.524m^3$, which gives us its volume $V=131.49m^3$. Then we apply the quadratic fitting function in \cite{diaz2005reverberation} to compute the reverberation time at $\SI{500}{\hertz}$ in furnished rooms as
\begin{equation}
\label{eq:rt500}
\text{T}_{60}=-2\times10^5V^2+0.0048V+0.255.
\end{equation}
This yields $\text{T}_{60}=0.54s$. Then we continue applying the regression equation in \cite{tang2004speech} to calculate an estimated STI as:
\begin{equation}
\label{eq:sti_est}
\text{STI}=0.5895-0.4422\text{log}_{10}(\text{T}_{60}),
\end{equation}
which gives $\text{STI}=0.708$. Then we use the same setup and computed IRs for five source-receiver pairs under a distribution in Figure~\ref{fig:sabine_setup} using both geometric propagation and hybrid propagation separately. Figure~\ref{fig:sabine} shows a comparison of STI values computed from our propagated IRs and the empirical model. 

\section{Conclusion and Future Work}
\label{sec:conclusion}

We present a novel optimization-based receiver placement algorithm to improve speech intelligibility. Using hybrid sound propagation, we can keep our performance costs low while maintaining accuracy at lower frequencies. We have applied our approach to complex indoor scenes and obtained considerable improvement in STI. To the best of our knowledge, this is the first algorithm that performs hybrid sound propagation optimization for this application. 

In the future, we would like to couple our method with existing denoising and dereverberation algorithms in addition to previous noise reduction techniques. For example, a computer design can be specified using reduced environmental noise and then improved STI. Then, dereverberation algorithms can be applied to the resulting signal.

The primary limitation of our technique is the quality of our input. We found that in general, our computed STI values tended to be higher than many measured STIs in similar environments. Although we modeled some environmental noise sources, designers using our algorithm would need to perform measurements of the ambient noise levels first. This is not straightforward to model when ambient noise or transmissive noise sources (e.g. a highway outside the building) are involved.

Additionally, in many cases, the directivity of human speech can influence the STI. Our method, however, is limited to omnidirectional sound sources.

\section*{Acknowledgements}
This research was supported by ARO grant W911NF14-1-0437 and NSF grant 1320644.


\bibliography{mybibfile}

\begin{thebibliography}{10}
\expandafter\ifx\csname url\endcsname\relax
  \def\url#1{\texttt{#1}}\fi
\expandafter\ifx\csname urlprefix\endcsname\relax\def\urlprefix{URL }\fi
\expandafter\ifx\csname href\endcsname\relax
  \def\href#1#2{#2} \def\path#1{#1}\fi

\bibitem{morales2016efficient}
N.~Morales, D.~Manocha, Efficient wave-based acoustic material design
  optimization, Computer-Aided Design 78 (2016) 83--92.

\bibitem{monks2000audioptimization}
M.~Monks, B.~M. Oh, J.~Dorsey, Audioptimization: Goal-based acoustic design,
  Computer Graphics and Applications, IEEE 20~(3) (2000) 76--90.

\bibitem{morales2018optimizing}
N.~Morales, D.~Manocha, Optimizing source placement for noise minimization
  using hybrid acoustic simulation, Computer-Aided Design 96 (2018) 1--12.

\bibitem{sriram2018robust}
A.~Sriram, H.~Jun, Y.~Gaur, S.~Satheesh, Robust speech recognition using
  generative adversarial networks, in: 2018 IEEE International Conference on
  Acoustics, Speech and Signal Processing (ICASSP), IEEE, 2018, pp. 5639--5643.

\bibitem{pallett2003look}
D.~S. Pallett, A look at nist's benchmark asr tests: past, present, and future,
  in: Automatic Speech Recognition and Understanding, 2003. ASRU'03. 2003 IEEE
  Workshop on, IEEE, 2003, pp. 483--488.

\bibitem{helander2014handbook}
M.~G. Helander, Handbook of human-computer interaction, Elsevier, 2014.

\bibitem{gillespie2002acoustic}
B.~W. Gillespie, L.~E. Atlas, Acoustic diversity for improved speech
  recognition in reverberant environments, in: Acoustics, Speech, and Signal
  Processing (ICASSP), 2002 IEEE International Conference on, Vol.~1, IEEE,
  2002, pp. I--557.

\bibitem{en1660268}
B.~EN, 60268-16: 2011.", Sound system equipment--Part 16: Objective rating of
  speech intelligibility by speech transmission index.

\bibitem{galster2007effect}
J.~A. Galster, The effect of room volume on speech recognition in enclosures
  with similar mean reverberation time, Ph.D. thesis (2007).

\bibitem{tashev2005reverberation}
I.~Tashev, D.~Allred, Reverberation reduction for improved speech recognition,
  Proc. Hands-Free Communication and Microphone Arrays.

\bibitem{feng2014speech}
X.~Feng, Y.~Zhang, J.~Glass, Speech feature denoising and dereverberation via
  deep autoencoders for noisy reverberant speech recognition, in: Acoustics,
  Speech and Signal Processing (ICASSP), 2014 IEEE International Conference on,
  IEEE, 2014, pp. 1759--1763.

\bibitem{tchorz1999model}
J.~Tchorz, B.~Kollmeier, A model of auditory perception as front end for
  automatic speech recognition, The Journal of the Acoustical Society of
  America 106~(4) (1999) 2040--2050.

\bibitem{hirsch2000aurora}
H.-G. Hirsch, D.~Pearce, The aurora experimental framework for the performance
  evaluation of speech recognition systems under noisy conditions, in:
  ASR2000-Automatic Speech Recognition: Challenges for the new Millenium ISCA
  Tutorial and Research Workshop (ITRW), 2000.

\bibitem{barker2015third}
J.~Barker, R.~Marxer, E.~Vincent, S.~Watanabe, The third ‘chime’speech
  separation and recognition challenge: Dataset, task and baselines, in:
  Automatic Speech Recognition and Understanding (ASRU), 2015 IEEE Workshop on,
  IEEE, 2015, pp. 504--511.

\bibitem{kinoshita2016summary}
K.~Kinoshita, M.~Delcroix, S.~Gannot, E.~A. Habets, R.~Haeb-Umbach,
  W.~Kellermann, V.~Leutnant, R.~Maas, T.~Nakatani, B.~Raj, et~al., A summary
  of the reverb challenge: state-of-the-art and remaining challenges in
  reverberant speech processing research, EURASIP Journal on Advances in Signal
  Processing 2016~(1) (2016) 7.

\bibitem{ko2017study}
T.~Ko, V.~Peddinti, D.~Povey, M.~L. Seltzer, S.~Khudanpur, A study on data
  augmentation of reverberant speech for robust speech recognition, in:
  Acoustics, Speech and Signal Processing (ICASSP), 2017 IEEE International
  Conference on, IEEE, 2017, pp. 5220--5224.

\bibitem{palomaki2004binaural}
K.~J. Palom{\"a}ki, G.~J. Brown, D.~Wang, A binaural processor for missing data
  speech recognition in the presence of noise and small-room reverberation,
  Speech Communication 43~(4) (2004) 361--378.

\bibitem{chen2016large}
J.~Chen, Y.~Wang, S.~E. Yoho, D.~Wang, E.~W. Healy, Large-scale training to
  increase speech intelligibility for hearing-impaired listeners in novel
  noises, The Journal of the Acoustical Society of America 139~(5) (2016)
  2604--2612.

\bibitem{saksela2015optimization}
K.~Saksela, J.~Botts, L.~Savioja, Optimization of absorption placement using
  geometrical acoustic models and least squares, The Journal of the Acoustical
  Society of America 137~(4) (2015) EL274--EL280.

\bibitem{robinson2014concert}
P.~W. Robinson, S.~Siltanen, T.~Lokki, L.~Savioja, Concert hall geometry
  optimization with parametric modeling tools and wave-based acoustic
  simulations, Building Acoustics 21~(1) (2014) 55--64.

\bibitem{khalilian2015joint}
H.~Khalilian, I.~V. Baji{\'c}, R.~G. Vaughan, Joint optimization of loudspeaker
  placement and radiation patterns for sound field reproduction, in: Acoustics,
  Speech and Signal Processing (ICASSP), 2015 IEEE International Conference on,
  IEEE, 2015, pp. 519--523.

\bibitem{d1997room}
P.~D'Antonio, T.~J. Cox, Room optimizer: A computer program to optimize the
  placement of listener, loudspeakers, acoustical surface treatment and room
  dimensions in critical listening rooms, in: Audio Engineering Society
  Convention 103, Audio Engineering Society, 1997.

\bibitem{houtgast2002past}
T.~Houtgast, H.~Steeneken, W.~Ahnert, L.~Braida, R.~Drullman, J.~Festen,
  K.~Jacob, P.~Mapp, S.~McManus, K.~Payton, et~al., Past, present and future of
  the Speech Transmission Index, Soesterberg: TNO, 2002.

\bibitem{wijngaarden1999objective}
S.~J.~v. Wijngaarden, H.~J. Steeneken, Objective prediction of speech
  intelligibility at high ambient noise levels using the speech transmission
  index, in: Sixth European Conference on Speech Communication and Technology,
  1999.

\bibitem{houtgast1984multi}
T.~Houtgast, H.~J.~M. Steeneken, A multi-language evaluation of the
  rasti-method for estimating speech intelligibility in auditoria, Acta
  Acustica united with Acustica 54~(4) (1984) 185--199.

\bibitem{houtgast1985review}
T.~Houtgast, H.~J. Steeneken, A review of the mtf concept in room acoustics and
  its use for estimating speech intelligibility in auditoria, The Journal of
  the Acoustical Society of America 77~(3) (1985) 1069--1077.

\bibitem{schroeder1981modulation}
M.~R. Schroeder, Modulation transfer functions: Definition and measurement,
  Acta Acustica united with Acustica 49~(3) (1981) 179--182.

\bibitem{cabrera2018critical}
D.~Cabrera, M.~Yadav, D.~Protheroe, Critical methodological assessment of the
  distraction distance used for evaluating room acoustic quality of open-plan
  offices, Applied Acoustics 140 (2018) 132--142.

\bibitem{cabrera2014increasing}
D.~Cabrera, D.~Lee, G.~Leembruggen, D.~Jimenez, Increasing robustness in the
  calculation of the speech transmission index from impulse responses, Building
  Acoustics 21~(3) (2014) 181--198.

\bibitem{raghuvanshi2009efficient}
N.~Raghuvanshi, R.~Narain, M.~C. Lin, Efficient and accurate sound propagation
  using adaptive rectangular decomposition, Visualization and Computer
  Graphics, IEEE Transactions on 15~(5) (2009) 789--801.

\bibitem{morales2015parallel}
N.~Morales, R.~Mehra, D.~Manocha, A parallel time-domain wave simulator based
  on rectangular decomposition for distributed memory architectures, Applied
  Acoustics 97 (2015) 104--114.

\bibitem{schissler2016interactive}
C.~Schissler, D.~Manocha, Interactive sound propagation and rendering for large
  multi-source scenes, ACM Transactions on Graphics (TOG) 36~(1) (2016) 2.

\bibitem{linkwitz1976active}
S.~H. Linkwitz, Active crossover networks for noncoincident drivers, Journal of
  the Audio Engineering Society 24~(1) (1976) 2--8.

\bibitem{bradley1999just}
J.~Bradley, R.~Reich, S.~Norcross, A just noticeable difference in c 50 for
  speech, Applied Acoustics 58~(2) (1999) 99--108.

\bibitem{egan2007architectural}
M.~D. Egan, Architectural Acoustics (J. Ross Publishing Classics), Vol.~4, J.
  Ross Publishing, 2007.

\bibitem{schissler2018acoustic}
C.~Schissler, C.~Loftin, D.~Manocha, Acoustic classification and optimization
  for multi-modal rendering of real-world scenes, IEEE transactions on
  visualization and computer graphics 24~(3) (2018) 1246--1259.

\bibitem{schissler2017interactive}
C.~Schissler, D.~Manocha, Interactive sound propagation and rendering for large
  multi-source scenes, ACM Transactions on Graphics (TOG) 36~(1) (2017) 2.

\bibitem{mehra2014acoustic}
R.~Mehra, N.~Raghuvanshi, A.~Chandak, D.~G. Albert, D.~Keith~Wilson,
  D.~Manocha, Acoustic pulse propagation in an urban environment using a
  three-dimensional numerical simulation, The Journal of the Acoustical Society
  of America 135~(6) (2014) 3231--3242.

\bibitem{diaz2005reverberation}
C.~D{\'\i}az, A.~Pedrero, The reverberation time of furnished rooms in
  dwellings, Applied Acoustics 66~(8) (2005) 945--956.

\bibitem{tang2004speech}
S.~Tang, M.~Yeung, Speech transmission index or rapid speech transmission index
  for classrooms? a designer's point of view, Journal of sound and vibration.

\end{thebibliography}

\end{document}